# Hardware Implementation of Double Pendulum Pseudo Random Number Generator

Jarrod Lim, Tom Manuel Opalla Piccio, Chua Min Jie Michelle, Maoyang Xiang, T. Hui Teo*
Engineering Product Development
Singapore University of Technology and Design
tthui@sutd.edu.sg

**ABSTRACT**

The objective of this project is to utilize an FPGA board which is the CMOD A7 35t to obtain a pseudo random number which can be used for encryption. We aim to achieve this by leveraging the inherent randomness present in environmental data captured by sensors. This data will be used as a seed to initialize an algorithm implemented on the CMOD A7 35t FPGA board. The project will focus on interfacing the sensors with the FPGA and developing suitable algorithms to ensure the generated numbers exhibit strong randomness properties.

# Introduction

This work aims to design a pseudo random number generator (pRNG) using double pendulum. Unlike the commonly seen single pendulum, the double pendulum is categorized as a dynamic chaotic system [1]. The double pendulum is modelled with two-points masses at the light rods, Figure 1. The environmental data are used to set the conditions (pRNG seed) such as weight, length, and gravitational acceleration onto a double pendulum (chaotic pendulum) algorithm to generate the random numbers. Therefore, the paper will be split into three components:

- Hardware and software interface for obtaining seed information from environment with sensors (Seed Generator),
- Capturing and displaying random numbers after user input, onto a LED screen,
- Double pendulum random number generator algorithm (Random Number Generator).

---

*Corresponding Author.
These authors contributed equally to this work.

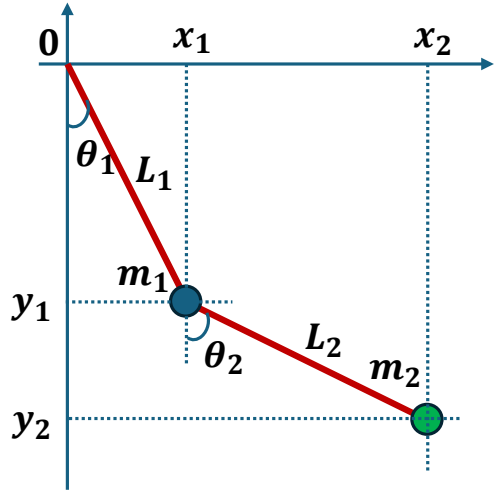

Figure 1: Double pendulum free body diagram.

## Pseudo Random Number Generator Algorithm

The pRNG algorithm is derived based on the chaotic pendulum. A chaotic pendulum forms a system that exhibits a rich dynamic behavior with strong sensitivity to initial conditions. These initial conditions are comprised of the initial position, lengths, and weights of the pendulum. These form the seed for the random number generator.

These values are then applied to the governing equations that describing the double pendulum [1].

$$\omega'_1 = \frac{-g(2m_1 + m_2)\sin\theta_1 - m_2 g \sin(\theta_1 - 2\theta_2) - 2\sin(\theta_1 - \theta_2)m_2[\omega_2^2 L_2 + \omega_1^2 L_1 \cos(\theta_1 - \theta_2)]}{L_1(2m_1 + m_2 - m_2 \cos(2\theta_1 - 2\theta_2))} \quad (1)$$

$$\omega'_2 = \frac{2\sin(\theta_1 - \theta_2)[\omega_1^2 L_1(m_1 + m_2) + g(m_1 + m_2)\cos\theta_1 + \omega_2^2 L_2 m_2 \cos(\theta_1 - \theta_2)]}{L_2(2m_1 + m_2 - m_2 \cos(2\theta_1 - 2\theta_2))} \quad (2)$$

where the notations that are referred to Fig. 1, Eq. (1), and Eq. (2):
$\omega_1 = \theta'_1$, $\omega_2 = \theta'_2$
$\omega'_1 = \theta''_1$, $\omega'_2 = \theta''_2$

$x \stackrel{\text{def}}{=}$ horizontal position of pendulum mass
$y \stackrel{\text{def}}{=}$ vertical position of pendulum mass
$\theta \stackrel{\text{def}}{=}$ angle of pendulum (0 = vertical downwards, counterclockwise is positive)
$L \stackrel{\text{def}}{=}$ length of rod (constant)
$T \stackrel{\text{def}}{=}$ tension in the rod
$m \stackrel{\text{def}}{=}$ mass of pendulum
$g \stackrel{\text{def}}{=}$ gravitational constant

The above equations (1) (2) are translated to the Verilog HDL code and implemented in the FPGA. However, the implementation of this algorithm is not straight forward as there are several issues. Firstly, the algorithm uses complex math with trigonometric functions. Secondly, values obtained when performing the functions are signed integers with decimal values. Verilog HDL is unable to perform these types of operations, hence functions must be developed to handle these exceptions.

*Corresponding Author.
These authors contributed equally to this work.

# Mathematic Functions

In order to handle the decimal number operation, functions for all basic operations such as plus, minus, times and divide are designed. The Verilog HDL code for this module is attached in Appendix. A 32 bits number system is used for this purpose. The first bit determines the sign of the value. The next 8 bits are the integer values while the last 23 bits control the decimal points (coded for 2 decimal place accuracy).

| +/- | Integer | | | | | | | | Decimal | | | | | | | | | | | | | | | | | | | | | | | |
|---|---|---|---|---|---|---|---|---|---|---|---|---|---|---|---|---|---|---|---|---|---|---|---|---|---|---|---|---|---|---|---|---|
| 31 | 30 | 29 | 28 | 27 | 26 | 25 | 24 | 23 | 22 | 21 | 20 | 19 | 18 | 17 | 16 | 15 | 14 | 13 | 12 | 11 | 10 | 9 | 8 | 7 | 6 | 5 | 4 | 3 | 2 | 1 | 0 |
| 0 | 0 | 0 | 1 | 0 | 0 | 0 | 0 | 1 | 0 | 0 | 0 | 0 | 0 | 0 | 0 | 0 | 0 | 0 | 0 | 0 | 0 | 0 | 0 | 0 | 0 | 1 | 0 | 0 | 1 | 0 | 0 | 1 |
| + | 33 | | | | | | | | 73 | | | | | | | | | | | | | | | | | | | | | | |
| +33.73 | | | | | | | | | | | | | | | | | | | | | | | | | | | | | | | |

Figure 2: Mapping of a 32 bits number system.

## MATH.plus Function

To perform a simple plus function of A.a + B.b. There are 4 key conditions to check to determine how the plus operation should be carried out in 16 different ways ($2^4$):
1. Sign of the number A
2. Sign of the number B
   a. E.g. If (+)A and (-)B, the plus operation becomes a minus
3. Which integer in the operation is larger
   a. If it is in fact a minus operation, if (-)B > (+)A, the resulting sign would flip
   b. However if (+)A > (-)B, the sign would remain.
4. Which decimal in the operation is larger
   a. Should A = B, we then must compare a and b. These add another layer of complexity apart from that seen in part 3. As there would be overflowing to create balance carried over to the integer.
   b. Or there could be insufficient values to perform the operation, hence we need to drop an integer to perform the decimal operation.

Please refer to the MATH function in the GitHub page for detailed explanation of the code.
The following is an example of a plus operation:

| | | +/- | Integer | | | | | | | | Decimal | | | | | | | | | | | | | | | | | | | | | | | |
|---|---|---|---|---|---|---|---|---|---|---|---|---|---|---|---|---|---|---|---|---|---|---|---|---|---|---|---|---|---|---|---|---|---|---|
| | | | | | | | | | | MATH.plus(A,B); | | | | | | | | | | | | | | | | | | | | | | | | |
| | | 31 | 30 | 29 | 28 | 27 | 26 | 25 | 24 | 23 | 22 | 21 | 20 | 19 | 18 | 17 | 16 | 15 | 14 | 13 | 12 | 11 | 10 | 9 | 8 | 7 | 6 | 5 | 4 | 3 | 2 | 1 | 0 | |
| A | | 0 | 0 | 0 | 1 | 0 | 0 | 0 | 0 | 1 | 0 | 0 | 0 | 0 | 0 | 0 | 0 | 0 | 0 | 0 | 0 | 0 | 0 | 0 | 0 | 0 | 1 | 0 | 0 | 1 | 0 | 0 | 1 | |
| | + | 33 | | | | | | | | 73 | | | | | | | | | | | | | | | | | | | | | | | | |
| B | | 1 | 0 | 0 | 0 | 1 | 0 | 1 | 0 | 1 | 0 | 0 | 0 | 0 | 0 | 0 | 0 | 0 | 0 | 0 | 0 | 0 | 0 | 0 | 0 | 1 | 0 | 1 | 0 | 1 | 0 | 0 | | |
| | - | 21 | | | | | | | | 84 | | | | | | | | | | | | | | | | | | | | | | | | |
| Z | | 0 | 0 | 0 | 0 | 0 | 1 | 1 | 0 | 0 | 0 | 0 | 0 | 0 | 0 | 0 | 0 | 0 | 0 | 0 | 0 | 0 | 0 | 0 | 0 | 0 | 0 | 1 | 0 | 1 | 1 | b > a, perform b - a |
| | + | 12 | | | | | | | | (-)11 | | | | | | | | | | | | | | | | | | | | | | Overflow | |
| Z | | 0 | 0 | 0 | 0 | 0 | 1 | 0 | 1 | 1 | 0 | 0 | 0 | 0 | 0 | 0 | 0 | 0 | 0 | 0 | 0 | 0 | 0 | 0 | 0 | 0 | 1 | 0 | 1 | 1 | 0 | 0 | 1 | Take from integers |
| | + | 11 | | | | | | | | 89 | | | | | | | | | | | | | | | | | | | | | | 100 - 11 | |
| +11.89 | | | | | | | | | | | | | | | | | | | | | | | | | | | | | | | | | | |

Figure 3: 32-bit Math.plus function.

*Corresponding Author.
These authors contributed equally to this work.

## MATH.minus Function

After developing the MATH.plus function, the minus function is a simple negation of the plus. I.e. (+)A - (+)B would be (+)A + (-)B. Hence to simplify the code, the MATH.plus function is called with the signed bit flipped. This is basically the example as seen above.

## MATH.times Function

The multiplication function is not as complicated as the plus or minus as we can simply implement a XOR gate on the signed bit and multiply the integers and decimals separately. Given the large value that can be obtained when doing multiplication on the decimals, more bits are allocated to the decimals.

The following mathematical manipulation is carried out to simplify the logic.

$$A.a * B.b = (A * B) + (A * .b) + (B * .a) + (.a + .b) \qquad (3)$$

where
- (A * B) outputs an integer
- (A * .b) and (B * .a) outputs a decimal overflow number
- (.a * .b) outputs pure decimals, with LHS digit corresponding to 1's place.

Sample multiplication for MATH.times(+A.a,-B.b):

| | +/- | Integer | | | | | | | | | Decimal | | | | | | | | | | | | | | | | | | | | | | | |
|---|---|---|---|---|---|---|---|---|---|---|---|---|---|---|---|---|---|---|---|---|---|---|---|---|---|---|---|---|---|---|---|---|---|---|
| | 31 | 30 | 29 | 28 | 27 | 26 | 25 | 24 | 23 | 22 | 21 | 20 | 19 | 18 | 17 | 16 | 15 | 14 | 13 | 12 | 11 | 10 | 9 | 8 | 7 | 6 | 5 | 4 | 3 | 2 | 1 | 0 | | |
| A + | 0 | 0 | 0 | 0 | 0 | 1 | 1 | 0 | 1 | 0 | 0 | 0 | 0 | 0 | 0 | 0 | 0 | 0 | 0 | 0 | 0 | 0 | 0 | 0 | 1 | 0 | 0 | 1 | 0 | 0 | 1 | | |
| | | | | | 13 | | | | | | | | | | | | | | 73 | | | | | | | | | | | | | | |
| B - | 1 | 0 | 0 | 0 | 0 | 0 | 1 | 1 | 1 | 0 | 0 | 0 | 0 | 0 | 0 | 0 | 0 | 0 | 0 | 0 | 0 | 0 | 0 | 0 | 1 | 0 | 1 | 0 | 1 | 0 | 0 | | |
| | | | | | 7 | | | | | | | | | | | | | | 84 | | | | | | | | | | | | | | |
| Z 1 - | | 0 | 0 | 0 | 0 | 1 | 1 | 0 | 0 | | | | | | | | | | | | | | | | | | | | | | | | | |
| | | | | A * B 91 | | | | | | A * .b (decimal overflow) 1092 | | | | | | B * .a (decimal overflow) 511 1664 | | | | | | | .a * .b (decimal only, 2dp) 6132 | | | | | | | | | | Overflow by 16 |
| Z - | 1 | 0 | 1 | 1 | 0 | 1 | 0 | 1 | 1 | 0 | 0 | 0 | 0 | 0 | 0 | 0 | 0 | 0 | 0 | 0 | 0 | 0 | 0 | 0 | 1 | 0 | 0 | 0 | 0 | 0 | 0 | | Add 16 to integer Take last 2 digits |
| | | | | | 107 | | | | | | | | | | | | | | 64 | | | | | | | | | | | | | | |
| | -107.64 | | | | | | | | | | | | | | | | | | | | | | | | | | | | | | | | |

Figure 4: 32-bit Math.times function.

## MATH.divide Function

What you notice when doing division is that regardless of the location of the decimal point, the resultant division is the same. Given that knowledge, for this function, the integers are combined with the decimal to create one larger number. From there modulus division with the onboard vision is done to obtain the quotient, from which we can derive the remainder. View the example below.

*Corresponding Author.
These authors contributed equally to this work.

## Figure 5

| | +/- | Integer | | | | | | | | Decimal | | | | | | | | | | | | | | | | | | | | | | | | |
|---|---|---|---|---|---|---|---|---|---|---|---|---|---|---|---|---|---|---|---|---|---|---|---|---|---|---|---|---|---|---|---|---|---|
| | | 31 | 30 | 29 | 28 | 27 | 26 | 25 | 24 | 23 | 22 | 21 | 20 | 19 | 18 | 17 | 16 | 15 | 14 | 13 | 12 | 11 | 10 | 9 | 8 | 7 | 6 | 5 | 4 | 3 | 2 | 1 | 0 |

(Figure 5 shows the MATH.divide(A,B) function working with 32-bit values, including rows for A=9.25, B=2.56, Integer=3 (A%B), Quotient=157 (A-(3*B)), Quotient*100=15700, Decimal=61, Z=+3.61)

*Figure 5: 32-bit Math.divide function.*

## MATH.sin Function

To create a sine function, instead of using a lookup table, function approximations are used. This is less complicated to code and more accurate as compared to a lookup table.

The following equation, Eq. (4) is used to for sine approximation.

$$\sin x = \frac{16x(\pi - x)}{5\pi^{\wedge}2 - 4x(\pi - x)} \qquad (4)$$

The following graph demonstrates the accuracy against an actual sine function:

*Figure 6: Approximate sine curve.*

The red line is the sine curve while the 2 blue lines represent the function approximation for 0 to $\pi$ and $\pi$ to $2\pi$.

For these functions to be usable, the MATH.sin function also handles negative theta inputs and theta inputs greater than $2\pi$. By shifting theta to within the boundaries.

*Corresponding Author.
These authors contributed equally to this work.

## MATH.cos Function

Understanding that $\cos\phi = \sin\left(\phi - \frac{\pi}{2}\right)$. The MATH.cos function basically calls the sin function with a modified theta input based on the cos to sin conversion.

# Hardware Implementation

Hardware implementation of the double pendulum pRNG in FPGA is summarized in this section.

## Generating seeds

After creating our own double pendulum algorithm, we decided to implement the sensors required to act as the seeds for our pRNG. We had decided on using four different types of sensors and they are:
- Magnetic field sensor,
- Microphone Sensitivity sound sensor,
- Photodiode light sensor,
- Temperature and Humidity Sensor.

These data will then be sent to a Nano Arduino via UART communication for the random number to be displayed on an LCD 1602 display screen. The block diagram of the hardware architect is depicted in Figure 7.

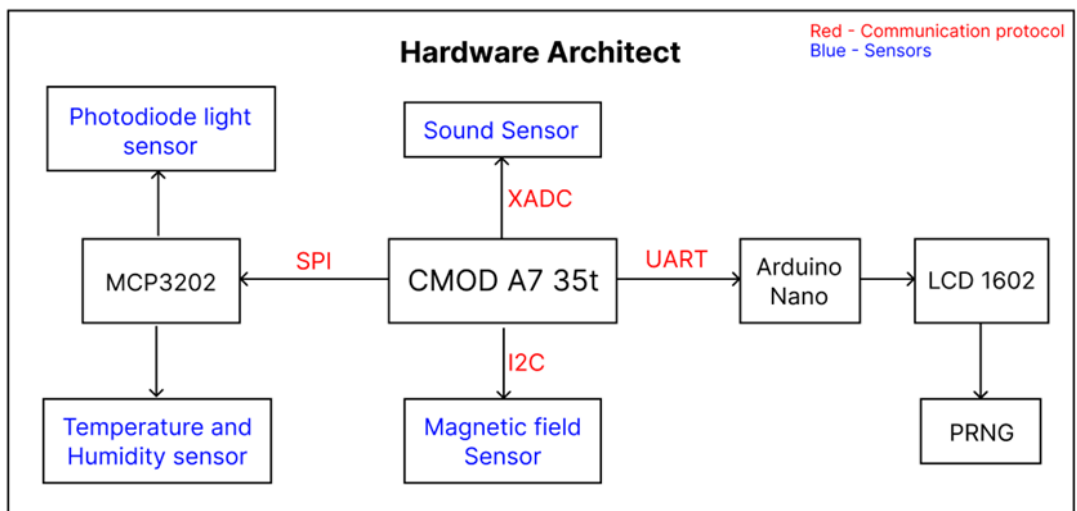

*Figure 7: Block diagram.*

## Magnetic Field Sensor

The HMC5883L or the Three-Axis Digital Compass IC does communication in Inter-Integrated Circuit (IIC or I2C) Communication. It is a special communication with a Serial Data Line (SDA) and Serial Clock Line (SCL), which are both inout nets. The magnetic hall sensors require 3 bytes to send and 7 bytes to receive. Hence, an FSM is designed to carry out the different states including the address of the registers and waiting to receive all 6 bytes of data (XYZ values of the compass with 16-bit signed number), [2].

However, a few values are changed to make it more random (uses only 1 sample instead of average of 8).

*Corresponding Author.
These authors contributed equally to this work.

- HMC5883L_ADDR = 7'h3C;
- CRA_VAL = 8'h10;   // Data: 0x10 0 00 100 00 - 1 sample at 15 Hz at normal configuration
- CRB_VAL = 8'h60;   // Data: 0x60 011 00000 Set gain to +-2.5
- MODE_VAL = 8'h01;  // Data: 0x01 means Single Mode
- READ_VAL = 8'h06;  //Data: Read all 6 data

The FSM of the I2C communication loops every one second and thus will have new data every second.

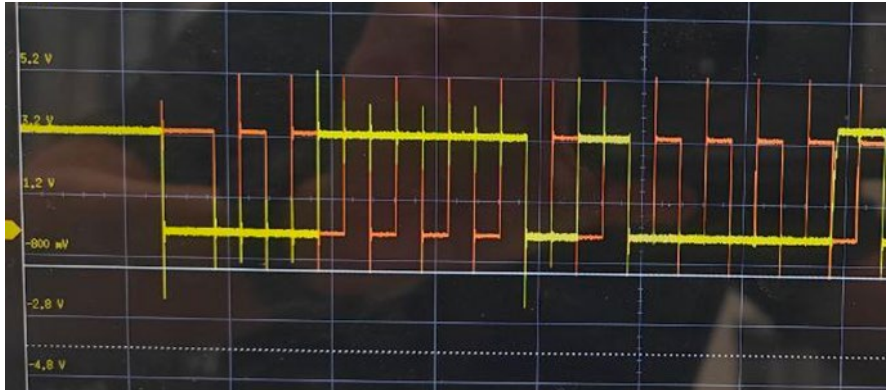

*Figure 8: Snapshot on oscilloscope displaying I2C's SDA (yellow) and SCL (orange) signals.*

As seen above, the FPGA is able to generate the I2C communication module and we were able to get non-zero values which means we have successfully communicated with the sensor. The data will be passed to the Arduino using UART.

# Analogue to Digital Converter

## XADC

The Xilinx Analog-to-Digital Converter (XADC), also referred to as XADC, is located on pins 15 and 16 of the CMOD A7 35t. These two pins can be configured to operate in either single-ended or differential mode. However, for our specific requirements, we only require the XADC to function in single-ended mode. Pin 15 is directly connected to the microphone sound sensor, while the remaining sensors will be connected via the PMOD connector.

## External ADC (SPI)

The PMOD connectors on the CMOD are utilized for connecting the MCP3202. Two MCP3202 modules have been incorporated to enable the connection of up to four sensors with an external ADC capability. Consequently, rather than assigning a particular channel for the ADC function, a modular approach is adopted wherein the ADC function is executed across all available channels. Subsequently, within the top module, the desired data output can be selected by invoking the specific adc_mode.

*Corresponding Author.
These authors contributed equally to this work.

## CMOD A7 to Arduino

The Universal Asynchronous Receiver/Transmitter (UART) module receives and sends the data, which comprises only 8 bits. Given the requirement to transmit 64 bits of data, the FSM_LOOP was created to ensure the transmission of all 64 bits in successive 8-bit segments. Additionally, a byte_counter is established to facilitate the transmission of 8 bytes of data with each button press for subsequent reading by the Arduino.

## Display pRNG on LCD

The goal is to display the randomly generated numbers on a 1602 Liquid Crystal Display (LCD) module. This display is chosen instead of a 7-segment due to its ability to display numbers up to 16 digits per row which accommodates to our pRNG algorithm. In order to display the generated values on to the LCD display, it is connected through Arduino.

On the LCD module, the readings for each sensor is displayed on the first line of the screen and the output of the pRNG number is displayed on the second line. Below is a breakdown of the first line information.
        A: Magnetic field sensor
        B: Microphone Sensitivity sound sensor
        C: Photodiode light sensor
        D: Temperature and Humidity Sensor

# Prototype and Measured Results

An integrated system is developed that houses an FPGA, Arduino Nano, Magnetic Field Sensor, Light Sensor, Humidity Sensor, Microphone Sensor and an LCD display, as shown in Figure 9. This integrated system obtains seeds from the sensors, runs the pRNG algorithm and displays it for the user. Seeds can be obtained from the environment by pressing the white button on the top right as seen in Figure 9. The bottom right dial adjusts the brightness of the LCD display.

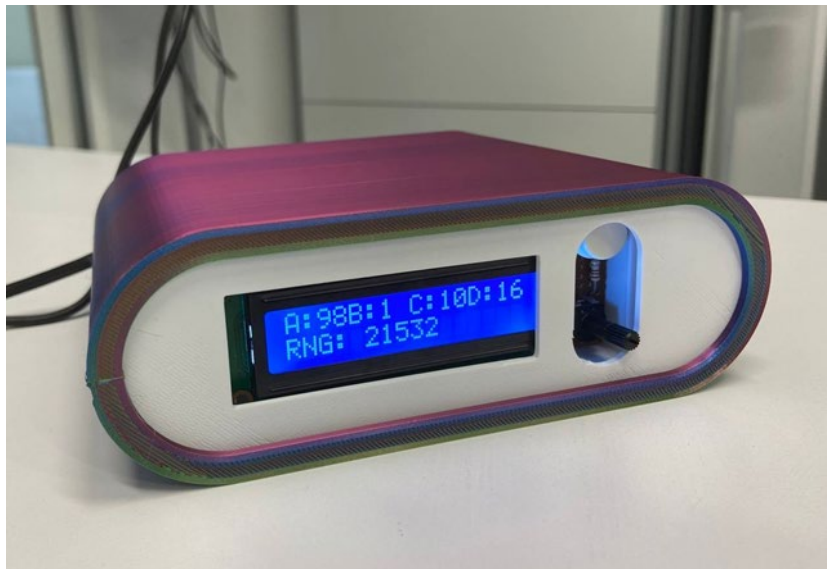
*Figure 9: pRNG prototype.*

*Corresponding Author.
These authors contributed equally to this work.

This pPRN generates a 10 digits number. The following is a histogram of 1,048,575 random numbers generated:

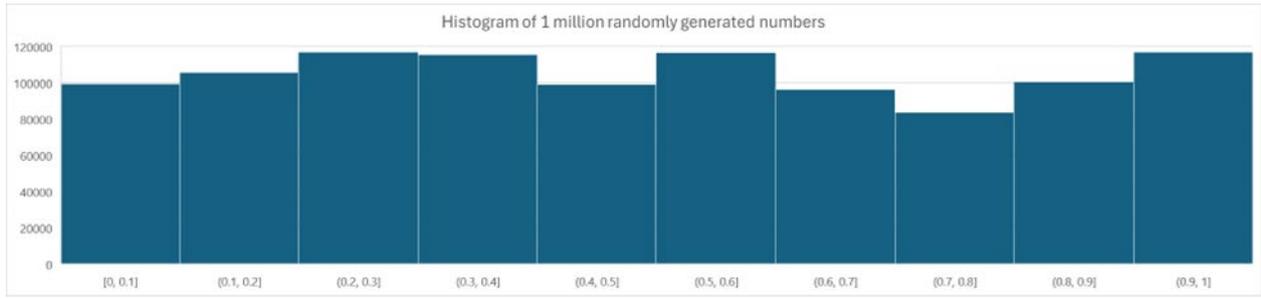
Figure 10: Histogram of recorded pRNG numbers.

As seen in this histogram, there is no discernible pattern in the histogram. A function written in Microsoft Excel also shows that the pattern does not repeat even after a million data points. Hence, the period for the random number generator does not repeat before a million cycles. There are over 1 billion possible combinations based on the 10 digits number.

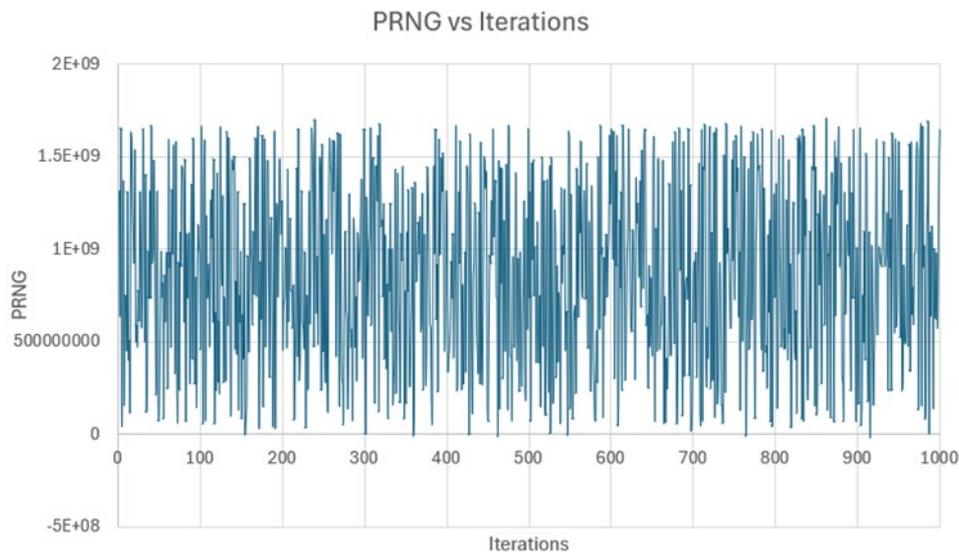
Figure 11: Plot of recorded pRNG number (first 1000 number).

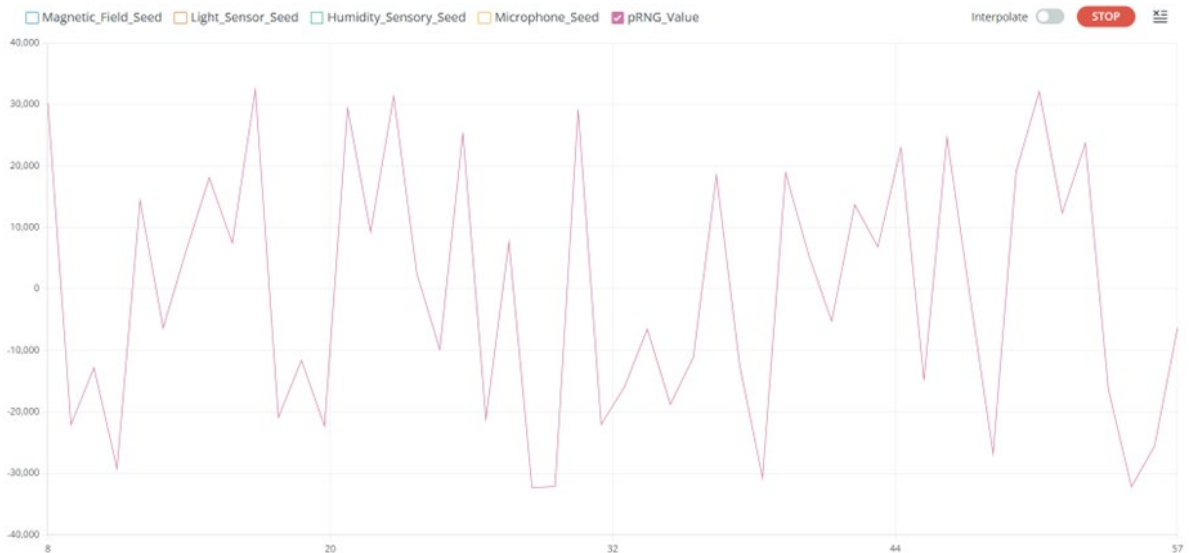
Figure 12: Live output plot.

*Corresponding Author.
These authors contributed equally to this work.

The two figures above plot the iterations versus time, Figure 11, Figure 12. These plots show that there is no repeat pattern of the pRNG number.

# Conclusions

The exercise showcases a double pendulum pRNG design and implementation in FPGA. The algorithm and hardware design can be further optimized for resource and speed. At the same time, more testing and evaluation are required to verify the pRNG performance.

# Acknowledgments

We would like to thank SUTD-ZJU IDEA Visiting Professor Grant (SUTD-ZJU (VP) 202103, and SUTD-ZJU Thematic Research Grant (SUTD-ZJU (TR) 202204), for supporting this work.

# Appendix

```
module Math(
    input [31:0]A, //1 pos/neg bit, 7 round bits, no decimals
    input [31:0]B //total of 8 bits
    );

function [31:0] plus;
    input [31:0] A; //bit 1 sign
    input [31:0] B; //bits 2 - 9 real
    reg [31:0]a, b; // bits 10 - 16 decimals
    reg [31:0]var; //local variable
    reg [8:0] symbol [1:0];
        begin
            symbol[1] = "-";
            symbol[0] = "+";
            a = A;
            b = B;
            case({a[31],b[31],~(a[30:23] >= b[30:23]), ~(a[22:0] >= b[22:0])})
                4'b0000, 4'b0001, 4'b0010, 4'b0011: begin //both positive
                    var[31] = 0;
                    var[30:23] = a[30:23] + b[30:23];
                    var[22:0] = a[22:0] + b[22:0];
                    while (var[22:0] >= 23'd100)begin
                        var[30:23] = var[30:23] + 8'd1;
                        var[22:0] = var[22:0] - 23'd100;
                        end
                    end
                    //----------------------------------------//
```

*Corresponding Author.
These authors contributed equally to this work.

```verilog
                4'b0100:begin //A positive, B negative, both of A larger than B
                        var[31] = 0;
                        var[30:23] = a[30:23] - b[30:23];
                        var[22:0] = a[22:0] - b[22:0];
                        end
                4'b0101:begin //A > B, a < b
                        var[31] = 0;
                        if(a[30:23] == b[30:23])begin
                            var[31] = 1;
                            a[30:23] = a[30:23]+ 8'd1;
                            var[22:0] = b[22:0] - a[22:0];end
                        var[30:23] = a[30:23] - b[30:23] - 8'd1;
                        var[22:0] = (a[22:0] + 23'd100) - b[22:0];
                        end
                4'b0110:begin //A < B, a > b
                        var[31] = 1;
                        if(a[30:23] == b[30:23])begin
                            b[30:23] = b[30:23]+ 8'd1; end
                        var[30:23] = b[30:23] - a[30:23] - 8'd1;
                        var[22:0] = (23'd100 + b[22:0]) - a[22:0];
                        end
                4'b0111:begin //A < B, a < b
                        var[31] = 1;
                        var[30:23] = b[30:23] - a[30:23];
                        var[22:0] = b[22:0] - a[22:0];
                        end
                        //---------------------------------------//
                4'b1000:begin //A negative, B positiv, both of A larger than B
                        var[31] = 1;
                        var[30:23] = a[30:23] - b[30:23];
                        var[22:0] = a[22:0] - b[22:0];
                        end
                4'b1001:begin //A > B, a < b
                        var[31] = 1;
                        if(a[30:23] == b[30:23])begin
                            a[30:23] = a[30:23]+ 8'd1; end
                        var[30:23] = a[30:23] - b[30:23] - 8'd1;
                        var[22:0] = (23'd100 + a[22:0]) - b[22:0];
                        end
                4'b1010:begin //A < B, a > b
                        var[31] = 0;
                        if(a[30:23] == b[30:23])begin
                            b[30:23] = b[30:23]+ 8'd1; end
                        var[30:23] = b[30:23] - a[30:23] - 8'd1;
                        var[22:0] = (23'd100 + b[22:0]) - a[22:0];
                        end
                4'b1011:begin //A < B, a < b
                        var[31] = 0;
                        var[30:23] = b[30:23] - a[30:23];
                        var[22:0] = b[22:0] - a[22:0];
                        end
                4'b1100, 4'b1101, 4'b1110, 4'b1111: begin
                    var[31] = 1;
                    var[30:23] = a[30:23] + b[30:23];
                    var[22:0] = a[22:0] + b[22:0];
                    while (var[22:0] >= 23'd100)begin
                        var[30:23] = var[30:23] + 8'd1;
                        var[22:0] = var[22:0] - 23'd100;
                        end
                    end
                default: var = 32'd100;
            endcase
            //$display("A = %c%d.%d, and , B = %c%d.%d",
symbol[A[31]],A[30:23],A[22:0],symbol[B[31]],B[30:23],B[22:0]);
            plus = var;
            //$display("A + B = %c%d.%d", symbol[plus[31]],plus[30:23],plus[22:0]);
//          $display("%c%d.%d + %c%d.%d = %c%d.%d"
//          ,symbol[a[31]],a[30:23],a[22:0]
//          ,symbol[b[31]],b[30:23],b[22:0]
//          ,symbol[plus[31]],plus[30:23],plus[22:0]);
```
11

*Corresponding Author.
These authors contributed equally to this work.

```verilog
        end
endfunction
//=======================================================================//
function [31:0] minus;
    input [31:0] A; //bit 1 sign
    input [31:0] B; //bits 2 - 9 real
    reg [31:0]var; //local variable
    reg [8:0] symbol [1:0];
    reg [31:0]a, b; // bits 10 - 16 decimals
        begin
            symbol[1] = "-";
            symbol[0] = "+";
            a = A;
            b = B;
            b[31] = ~B[31];
            var = plus(a,b);
            //$display("A = %c%d.%d, and , B = %c%d.%d",
symbol[A[31]],A[30:23],A[22:0],symbol[B[31]],B[30:23],B[22:0]);
            minus = var;
            //$display("A - B = %c%d.%d", symbol[minus[31]],minus[30:23],minus[22:0]);
        end
endfunction
//=======================================================================//
function [31:0] times;
    input [31:0] A; //bit 1 sign
    input [31:0] B; //bits 2 - 9 real
    reg [31:0]a, b; // bits 10 - 16 decimals
    reg [31:0]var, var1, var2, var3, var4, var5, dec;
    reg [8:0] symbol [1:0]; //(A.a)(B.b) = (A + .a)(B + .b) = (A*B + A*.b + B*.a + .a*.b)
        begin
            symbol[1] = "-";
            symbol[0] = "+";
            a = A;
            b = B;
            var1[30:23] = a[30:23] + b[30:23]; //A*B
            var2[22:0] = a[30:23] * b[22:0];    //A * .b
            var3[22:0] = b[30:23] * a[22:0];    //B * .a
            var4[22:0] = a[22:0] * b[22:0];     //.a * .b
            //first add the decimals together
            //Add first 2 DIGITS of var4 into var2 + var3, store in var5
            //extract first 2 digits of var4
            var5[7:0] = var4[22:0]/23'd100;
            dec[22:0] = var2[22:0] + var3[22:0] + var5[7:0];
            //next overflow the decimals into the integers
            var5[22:0] = dec[22:0]/23'd100;
            var[30:23] = (a[30:23] * b[30:23]) + var5[7:0];
            //store remaining decimals
            var[22:0] = dec[22:0] - (dec[22:0]/23'd100)*23'd100; //purposely truncated
            case({a[31],b[31] })
                2'b00, 2'b11: var[31] = 0; //both positive
                2'b01, 2'b10: var[31] = 1;
                default: var = 32'd100;
            endcase
            //$display("dec[22:0] = %d, dec[22:0]/100 = %d", dec[22:0], dec[22:0]/23'd100);
            //$display("A = %c%d.%d, and , B = %c%d.%d",
symbol[A[31]],A[30:23],A[22:0],symbol[B[31]],B[30:23],B[22:0]);
            times = var;
            //$display("A + B = %c%d.%d", symbol[times[31]],times[30:23],times[22:0]);
//          $display("%c%d.%d * %c%d.%d = %c%d.%d"
//              ,symbol[a[31]],a[30:23],a[22:0]
//              ,symbol[b[31]],b[30:23],b[22:0]
//              ,symbol[times[31]],times[30:23],times[22:0]);
        end
endfunction
//=======================================================================//
function [31:0] divide;
    input [31:0] A; //bit 1 sign
    input [31:0] B; //bits 2 - 9 real
    reg [31:0]a, b; // bits 10 - 16 decimals
    reg [31:0]var, int, quo, dec;
```

*Corresponding Author.
These authors contributed equally to this work.

```verilog
        reg [8:0] symbol [1:0]; //(A.a)/(B.b) ==> (Aa)(Bb) remove decimals and do division, keeping quotients
        begin
            symbol[1] = "-";
            symbol[0] = "+";
            a = A;
            b = B;
            //move decimal place and combine
            a[30:0] = A[30:23] * 8'd100;
            a[30:0] = a[30:0] + A[7:0];
            b[30:0] = B[30:23] * 8'd100;
            b[30:0] = b[30:0] + B[7:0];
//              $display("A = %c%d.%d, a = %c%d"
//              ,symbol[A[31]],A[30:23],A[22:0]
//              ,symbol[a[31]],a[30:0]);
            case({a[31],b[31],~(a[30:0] >= b[30:0])})
                3'b000, 3'b110: begin
                    var[31] = 0;
                    int[30:23] = a[30:0] / b[30:0];
                    quo[30:0] = a[30:0] - (int[30:23] * b[30:0]); //quotient
                    dec[22:0] = (quo[30:0]*100) / b[30:0]; //2dp is enough
                    var[30:23] = int[30:23];
                    var[22:0] = dec[22:0];end
                3'b001, 3'b111: begin
                    var[31] = 0;
                    var[30:23] = 0;
                    var[22:0] = (a[30:0] * 100) / b[30:0];end
                3'b011, 3'b101: begin
                    var[31] = 1;
                    var[30:23] = 0;
                    var[22:0] = (a[30:0] * 100) / b[30:0];end
                3'b010, 3'b100: begin
                    var[31] = 1;
                    int[30:23] = a[30:0] / b[30:0];
                    quo[30:0] = a[30:0] - (int[30:23] * b[30:0]); //quotient
                    dec[22:0] = (quo[30:0]*100) / b[30:0]; //2dp is enough
                    var[30:23] = int[30:23];
                    var[22:0] = dec[22:0];end
                default: var = 32'd100;
            endcase
            //$display("a = %d.%d divide by b = %d.%d", A[30:23],A[22:0], B[30:23],B[22:0]);
            //$display("Ans = %d.%d", var[30:23], var[22:0]);
            //$display("A = %c%d.%d, and , B = %c%d.%d",
symbol[A[31]],A[30:23],A[22:0],symbol[B[31]],B[30:23],B[22:0]);
            divide = var;
//              $display("%c%d.%d / %c%d.%d = %c%d.%d"
//              ,symbol[a[31]],a[30:23],a[22:0]
//              ,symbol[b[31]],b[30:23],b[22:0]
//              ,symbol[divide[31]],divide[30:23],divide[22:0]);
        end
endfunction
//=======================================================================//
function [31:0] mod;
    input [31:0] A;
    reg [31:0]a;
    reg [31:0]var; //local variable
    reg [8:0] symbol [1:0];
        begin
            symbol[1] = "-";
            symbol[0] = "+";
            a = A;
            var[31] = 0;
            var[30:0] = a[30:0];
//              $display("A = %c%d.%d", symbol[A[31]],A[30:23],A[22:0]);
            mod = var;
//              $display("|A| = %c%d.%d", symbol[var[31]],mod[30:23],mod[22:0]);
        end
endfunction
//=======================================================================//
function [31:0] neg;
    input [31:0] A;
```

*Corresponding Author.
These authors contributed equally to this work.

```verilog
        reg [31:0]a;
        reg [31:0]var; //local variable
        reg [8:0] symbol [1:0];
            begin
                symbol[1] = "-";
                symbol[0] = "+";
                a = A;
                var[31] = ~a[31];
                var[30:0] = a[30:0];
//              $display("A = %c%d.%d", symbol[A[31]],A[30:23],A[22:0]);
                neg = var;
//              $display("neg A = %c%d.%d", symbol[var[31]],neg[30:23],neg[22:0]);
            end
endfunction
//========================================================================//
reg [31:0]pi, pi_2;
reg [31:0]const16, const5, const4, const2, const1;
initial begin
const16 = 0; const5 = 0; const4 = 0; const2 = 0; const1 = 0;
const16[30:23] = 8'd16;
const5[30:23] = 8'd5;
const4[30:23] = 8'd4;
const2[30:23] = 8'd2;
const1[30:23] = 8'd1;

pi[31] = 0;
pi[30:23] = 8'd3;
pi[22:0] = 23'd14;
pi_2 = Math.times(pi,const2);
end

function [31:0] sin;
    input [31:0]theta;
    reg [31:0] var, store, var1, var2, var3, var4, var5, var6, var7;
    reg [8:0] symbol [1:0];
    begin
    symbol[1] = "-";
    symbol[0] = "+";
    //Check what is theta
    store = divide(theta,pi_2);
    case({theta[31],store[30:23] >= 1})
        2'b00:begin//positive number less than 2 pi
        theta = theta;end
        2'b01:begin//positive numver greater than 2 pi
        store = divide(theta,pi_2);
        store[31] = 0; store[22:0] = 0;
        theta = minus(theta,times(store,pi_2));end
        2'b10:begin//negative number less than 2pi
        theta = mod(theta);
        theta = minus(pi_2,theta);end
        2'b11:begin//negatiev number greater than 2pi
        //$display("theta before = %c%d.%d", symbol[theta[31]],theta[30:23],theta[22:0]);
        store = 0; store = divide(theta,pi_2); store[31] = 0; store[22:0] = 23'd0;//1.11
        theta = plus(theta,times(plus(const1,store),pi_2)); //7-2pi
        //$display("theta after= %c%d.%d", symbol[theta[31]],theta[30:23],theta[22:0]);
        end
    endcase
    if(theta >= 0 && theta <= pi)begin //first 2 quadrants
        //sine function for 0 to pi:   y = 16x(pi-x) / (5pi^2 - 4x(pi-x))
        var = times(times(const16,theta),minus(pi,theta));
        var = divide(var,times(times(pi,pi),const5) - times(times(const4,theta),minus(pi,theta)));
        //$display("i went here");
        end
    if(theta > pi && theta <= pi_2)begin //last 2 quadrants ///ERROR HERE
        //sine function for pi to 2pi: y = -16(x-2pi)(pi-x) / (5pi^2 - 4(x-2pi)(pi-x))
        var = times(times(neg(const16), minus(theta,pi_2)),minus(pi,theta));
        //$display("var part 1 = %c%d.%d", symbol[var[31]],var[30:23],var[22:0]);
        var = divide(var,minus(times(times(pi,pi),const5),times(times(const4,minus(theta,pi_2)),minus(pi,theta))));
        //$display("var part 2 = %c%d.%d", symbol[var[31]],var[30:23],var[22:0]);
```

*Corresponding Author.
These authors contributed equally to this work.

```verilog
        end
    sin = var;
    //$display("pi = %c%d.%d", symbol[pi[31]],pi[30:23],pi[22:0]);
    //$display("theta = %c%d.%d", symbol[theta[31]],theta[30:23],theta[22:0]);
    //$display("sin(theta) = %c%d.%d", symbol[sin[31]],sin[30:23],sin[22:0]);
    end
endfunction
//===========================================================================//
function [31:0] cos;
    input [31:0]theta;
    reg [31:0] var, store;
    reg [8:0] symbol [1:0];
    begin
    symbol[1] = "-";
    symbol[0] = "+";
    //cos(x) = sin(x - pi/2)
    cos = sin(minus(divide(pi,const2),theta));
    //$display("pi = %c%d.%d", symbol[pi[31]],pi[30:23],pi[22:0]);
    //$display("theta = %c%d.%d", symbol[theta[31]],theta[30:23],theta[22:0]);
    //$display("cos(theta) = %c%d.%d", symbol[cos[31]],cos[30:23],cos[22:0]);
    end
endfunction
//===========================================================================//
endmodule
```

*Corresponding Author.
These authors contributed equally to this work.